\pdfoutput=1
\documentclass[12pt,a4paper]{article}

\usepackage{ifthen} 
\newboolean{pdflatex}
\setboolean{pdflatex}{true} 

\newboolean{articletitles}
\setboolean{articletitles}{true} 

\newboolean{uprightparticles}
\setboolean{uprightparticles}{false} 

\def\paperauthors{LHCb collaboration} 
\def\paperasciititle{Measurement of the ratio of branching fractions BR(Bc2BsPi)/BR(Bc2JpsiPi)} 
\def\papertitle{Measurement of the ratio of branching fractions $\BR(\Bcp \to \Bs \pip)/\BR(\Bcp \to \jpsi \pip)$} 
\def\paperkeywords{{High Energy Physics}, {LHCb}} 
\def\papercopyright{\the\year\ CERN for the benefit of the LHCb collaboration} 
\def\paperlicence{CC BY 4.0 licence}
\def\paperlicenceurl{https://creativecommons.org/licenses/by/4.0/}

\usepackage[top=1in, bottom=1.25in, left=1in, right=1in]{geometry}

\usepackage{microtype}
\usepackage{lineno}  
\usepackage{xspace} 
\usepackage{caption} 

\usepackage{graphicx}  
\usepackage{color}
\usepackage{colortbl}
\graphicspath{{./figs/}} 

\usepackage{amsmath} 
\usepackage{amssymb}
\usepackage{amsfonts}
\usepackage{upgreek} 

\newcommand*\patchAmsMathEnvironmentForLineno[1]{%
\expandafter\let\csname old#1\expandafter\endcsname\csname #1\endcsname
\expandafter\let\csname oldend#1\expandafter\endcsname\csname
end#1\endcsname
 \renewenvironment{#1}%
   {\linenomath\csname old#1\endcsname}%
   {\csname oldend#1\endcsname\endlinenomath}%
}
\newcommand*\patchBothAmsMathEnvironmentsForLineno[1]{%
  \patchAmsMathEnvironmentForLineno{#1}%
  \patchAmsMathEnvironmentForLineno{#1*}%
}
\AtBeginDocument{%
\patchBothAmsMathEnvironmentsForLineno{equation}%
\patchBothAmsMathEnvironmentsForLineno{align}%
\patchBothAmsMathEnvironmentsForLineno{flalign}%
\patchBothAmsMathEnvironmentsForLineno{alignat}%
\patchBothAmsMathEnvironmentsForLineno{gather}%
\patchBothAmsMathEnvironmentsForLineno{multline}%
\patchBothAmsMathEnvironmentsForLineno{eqnarray}%
}

\usepackage{hyperxmp}

\usepackage[pdftex,
            pdfauthor={\paperauthors},
            pdftitle={\paperasciititle},
            pdfkeywords={\paperkeywords},
            pdfcopyright={Copyright (C) \papercopyright},
            pdflicenseurl={\paperlicenceurl}]{hyperref}

\usepackage[colorinlistoftodos,textsize=scriptsize]{todonotes}

\usepackage[bottom,flushmargin,hang,multiple]{footmisc}

\usepackage[all]{hypcap} 

\usepackage{xspace} 
\usepackage{upgreek}

\def\lhcb   {\mbox{LHCb}\xspace}

\def\MagUp {\mbox{\em Mag\kern -0.05em Up}\xspace}

\ifthenelse{\boolean{uprightparticles}}%
{

 \def\Pmu         {\ensuremath{\upmu}\xspace}                 
 \def\Pnu         {\ensuremath{\upnu}\xspace}                 
                  
 \def\Ppi         {\ensuremath{\uppi}\xspace}

 \def\Ptau        {\ensuremath{\uptau}\xspace}                 
                  
 \def\Pphi        {\ensuremath{\upphi}\xspace}

 \def\Ppsi        {\ensuremath{\uppsi}\xspace}

 \def\PDelta      {\ensuremath{\Delta}\xspace}                 
 \def\PXi         {\ensuremath{\Xi}\xspace}                 
 \def\PLambda     {\ensuremath{\Lambda}\xspace}                 
 \def\PSigma      {\ensuremath{\Sigma}\xspace}                 
 \def\POmega      {\ensuremath{\Omega}\xspace}                 
 \def\PUpsilon    {\ensuremath{\Upsilon}\xspace}
 \let\oldPi\Pi
 \def\PPi         {\ensuremath{\oldPi}\xspace}

 \def\PB      {\ensuremath{\mathrm{B}}\xspace}                 
                  
 \def\PD      {\ensuremath{\mathrm{D}}\xspace}

 \def\PJ      {\ensuremath{\mathrm{J}}\xspace}                 
 \def\PK      {\ensuremath{\mathrm{K}}\xspace}

 \def\PW      {\ensuremath{\mathrm{W}}\xspace}

 \def\Pb      {\ensuremath{\mathrm{b}}\xspace}                 
 \def\Pc      {\ensuremath{\mathrm{c}}\xspace}

 \def\Pi      {\ensuremath{\mathrm{i}}\xspace}

 \def\Ps      {\ensuremath{\mathrm{s}}\xspace}

 \def\thebaroffset{0.0em}
}
{

 \def\Pmu         {\ensuremath{\mu}\xspace}                 
 \def\Pnu         {\ensuremath{\nu}\xspace}                 
                  
 \def\Ppi         {\ensuremath{\pi}\xspace}

 \def\Ptau        {\ensuremath{\tau}\xspace}                 
                  
 \def\Pphi        {\ensuremath{\phi}\xspace}

 \def\Ppsi        {\ensuremath{\psi}\xspace}                 
                  
 \mathchardef\PDelta="7101
 \mathchardef\PXi="7104
 \mathchardef\PLambda="7103
 \mathchardef\PSigma="7106
 \mathchardef\POmega="710A
 \mathchardef\PUpsilon="7107
 \mathchardef\PPi="7105
                  
 \def\PB      {\ensuremath{B}\xspace}                 
                  
 \def\PD      {\ensuremath{D}\xspace}

 \def\PJ      {\ensuremath{J}\xspace}                 
 \def\PK      {\ensuremath{K}\xspace}

 \def\PW      {\ensuremath{W}\xspace}

 \def\Pb      {\ensuremath{b}\xspace}                 
 \def\Pc      {\ensuremath{c}\xspace}

 \def\Pi      {\ensuremath{i}\xspace}

 \def\Ps      {\ensuremath{s}\xspace}

 \def\thebaroffset{0.18em}
}
\newcommand{\offsetoverline}[2][\thebaroffset]{\kern #1\overline{\kern -#1 #2}}%

\makeatletter
\ifcase \@ptsize \relax
  \newcommand{\miniscule}{\@setfontsize\miniscule{4}{5}}
\or
  \newcommand{\miniscule}{\@setfontsize\miniscule{5}{6}}
\or
  \newcommand{\miniscule}{\@setfontsize\miniscule{5}{6}}
\fi
\makeatother

\DeclareRobustCommand{\optbar}[1]{\shortstack{{\miniscule (\rule[.5ex]{1.25em}{.18mm})}
  \\ [-.7ex] $#1$}}

\def\mup        {{\ensuremath{\Pmu^+}}\xspace}

\def\mumu       {{\ensuremath{\Pmu^+\Pmu^-}}\xspace}

\def\taup       {{\ensuremath{\Ptau^+}}\xspace}

\def\neu        {{\ensuremath{\Pnu}}\xspace}

\def\neum       {{\ensuremath{\neu_\mu}}\xspace}

\def\neut       {{\ensuremath{\neu_\tau}}\xspace}

\def\Wp     {{\ensuremath{\PW^+}}\xspace}

\def\squark    {{\ensuremath{\Ps}}\xspace}

\def\cquark    {{\ensuremath{\Pc}}\xspace}

\def\bquark    {{\ensuremath{\Pb}}\xspace}
\def\bquarkbar {{\ensuremath{\overline \bquark}}\xspace}

\def\pion   {{\ensuremath{\Ppi}}\xspace}

\def\pip    {{\ensuremath{\pion^+}}\xspace}
\def\pim    {{\ensuremath{\pion^-}}\xspace}

\def\kaon    {{\ensuremath{\PK}}\xspace}

\def\KorKbar {\kern \thebaroffset\optbar{\kern -\thebaroffset \PK}{}\xspace}

\def\Kp      {{\ensuremath{\kaon^+}}\xspace}
\def\Km      {{\ensuremath{\kaon^-}}\xspace}

\newcommand{\phiz}{\ensuremath{\Pphi}\xspace}

\def\D       {{\ensuremath{\PD}}\xspace}

\def\DorDbar {\kern \thebaroffset\optbar{\kern -\thebaroffset \PD}\xspace}
\def\Dz      {{\ensuremath{\D^0}}\xspace}

\def\Dp      {{\ensuremath{\D^+}}\xspace}
\def\Dm      {{\ensuremath{\D^-}}\xspace}

\def\DpDm    {\ensuremath{\Dp {\kern -0.16em \Dm}}\xspace}

\def\Dstarp  {{\ensuremath{\D^{*+}}}\xspace}

\def\Dsm     {{\ensuremath{\D^-_\squark}}\xspace}

\def\B       {{\ensuremath{\PB}}\xspace}

\def\BorBbar {\kern \thebaroffset\optbar{\kern -\thebaroffset \PB}\xspace}

\def\Bd      {{\ensuremath{\B^0}}\xspace}

\def\BdorBdbar {\kern \thebaroffset\optbar{\kern -\thebaroffset \Bd}\xspace}

\def\Bs      {{\ensuremath{\B^0_\squark}}\xspace}

\def\BsorBsbar {\kern \thebaroffset\optbar{\kern -\thebaroffset \Bs}\xspace}
\def\Bc      {{\ensuremath{\B_\cquark^+}}\xspace}
\def\Bcp     {{\ensuremath{\B_\cquark^+}}\xspace}

\def\jpsi     {{\ensuremath{{\PJ\mskip -3mu/\mskip -2mu\Ppsi}}}\xspace}

\def\Y#1S{\ensuremath{\PUpsilon{(#1S)}}\xspace}

\def\LorLbar     {\kern \thebaroffset\optbar{\kern -\thebaroffset \PLambda}\xspace}

\def\BF         {{\ensuremath{\mathcal{B}}}\xspace}
\def\BR         {\BF}

\newcommand{\decay}[2]{\ensuremath{#1\!\to #2}\xspace} 

\def\to                 {\ensuremath{\rightarrow}\xspace}

\def\BsToJPsiPhi  {\decay{\Bs}{\jpsi\phi}}

\def\AT#1     {\ensuremath{A_{\mathrm{T}}^{#1}}\xspace}           

\def\C#1      {\ensuremath{\mathcal{C}_{#1}}\xspace}                       
\def\Cp#1     {\ensuremath{\mathcal{C}_{#1}^{'}}\xspace}                    
\def\Ceff#1   {\ensuremath{\mathcal{C}_{#1}^{\mathrm{(eff)}}}\xspace}        
\def\Cpeff#1  {\ensuremath{\mathcal{C}_{#1}^{'\mathrm{(eff)}}}\xspace}       
\def\Ope#1    {\ensuremath{\mathcal{O}_{#1}}\xspace}                       
\def\Opep#1   {\ensuremath{\mathcal{O}_{#1}^{'}}\xspace}                    

\newcommand{\nospaceunit}[1]{\ensuremath{\text{#1}}}       
\newcommand{\aunit}[1]{\ensuremath{\text{\,#1}}}       

\newcommand{\tev}{\aunit{Te\kern -0.1em V}\xspace}
\newcommand{\gev}{\aunit{Ge\kern -0.1em V}\xspace}
\newcommand{\mev}{\aunit{Me\kern -0.1em V}\xspace}
\newcommand{\kev}{\aunit{ke\kern -0.1em V}\xspace}
\newcommand{\ev}{\aunit{e\kern -0.1em V}\xspace}
 
\newcommand{\mevc}{\ensuremath{\aunit{Me\kern -0.1em V\!/}c}\xspace}
\newcommand{\gevc}{\ensuremath{\aunit{Ge\kern -0.1em V\!/}c}\xspace}
\newcommand{\mevcc}{\ensuremath{\aunit{Me\kern -0.1em V\!/}c^2}\xspace}
\newcommand{\gevcc}{\ensuremath{\aunit{Ge\kern -0.1em V\!/}c^2}\xspace}

\def\mum  {\ensuremath{\,\upmu\nospaceunit{m}}\xspace}

\def\fb   {\ensuremath{\aunit{fb}}\xspace}
\def\invfb   {\ensuremath{\fb^{-1}}\xspace}

\def\ps   {\ensuremath{\aunit{ps}}\xspace}

\newcommand{\stat}{\aunit{(stat)}\xspace}
\newcommand{\syst}{\aunit{(syst)}\xspace}

\def\gsim{{~\raise.15em\hbox{$>$}\kern-.85em
          \lower.35em\hbox{$\sim$}~}\xspace}
\def\lsim{{~\raise.15em\hbox{$<$}\kern-.85em
          \lower.35em\hbox{$\sim$}~}\xspace}

\def\pt         {\ensuremath{p_{\mathrm{T}}}\xspace}

\def\ptot       {\ensuremath{p}\xspace}

\def\bcvegpy    {\mbox{\textsc{Bcvegpy}}\xspace}

\def\evtgen     {\mbox{\textsc{EvtGen}}\xspace}

\def\geant      {\mbox{\textsc{Geant4}}\xspace}

\def\photos     {\mbox{\textsc{Photos}}\xspace}

\def\pythia     {\mbox{\textsc{Pythia}}\xspace}

\def\tell1  {TELL1\xspace}
\def\ukl1   {UKL1\xspace}

\newcommand{\lhcborcid}[1]{\href{https://orcid.org/#1}{\hspace*{0.1em}\raisebox{-0.45ex}{\includegraphics[width=1em]{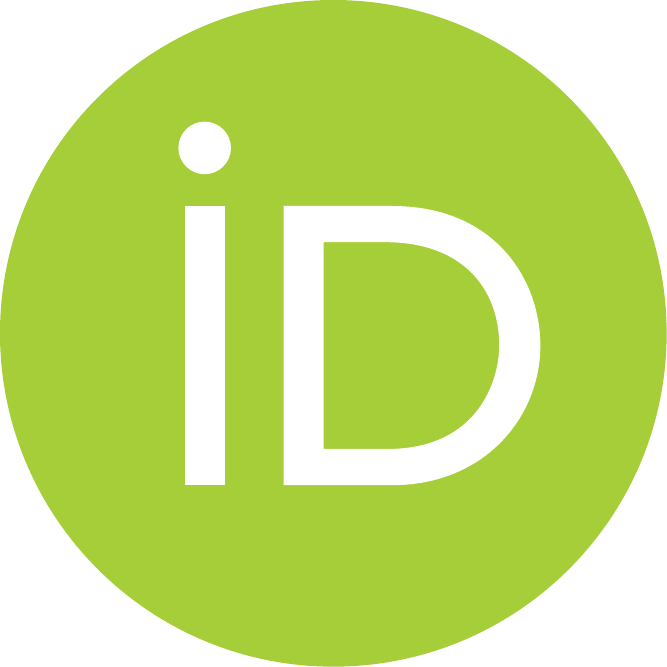}}}}

\usepackage{cite} 
\usepackage{mciteplus}

\newcommand{\bsubc}{\ensuremath{B_c^+}}

\newcommand{\JpsiPi}{\ensuremath{\bsubc \to \jpsi \pip}}

\newcommand{\BsPi}{\ensuremath{\bsubc \to \Bs \pip}}

\newcommand{\BsToDsPi}{\ensuremath{\Bs \to \Dsm \pip}}

\newcommand{\DsToKKPi}{\ensuremath{D_s^{-} \to K^+K^- \pim}}

\def\DsPi{D_s^-\pi^+}

\def\TOS{\texttt{TOS}}
\def\TIS{\texttt{TIS}}

\newcommand{\RJpsiPhi}{\ensuremath{\mathcal{R}_{\jpsi\phi}}}
\newcommand{\RDsPi}{\ensuremath{\mathcal{R}_{\DsPi}}}

\newcommand{\BsPiJpsiPhiArraw}{\ensuremath{\Bcp \to \Bs (\to \jpsi\phi) \pip}}
\newcommand{\BsPiDsPiArraw}{\ensuremath{\Bcp \to \Bs (\to \Dsm\pip) \pip}}

\newcommand{\BsPiJpsiPhi}{\BsPiJpsiPhiArraw}
\newcommand{\BsPiDsPi}{\BsPiDsPiArraw}

\usepackage{longtable} 
\usepackage{booktabs}
\usepackage{subfig}

\begin{document}

\renewcommand{\thefootnote}{\fnsymbol{footnote}}
\setcounter{footnote}{1}


\begin{titlepage}
\pagenumbering{roman}

\vspace*{-1.5cm}
\centerline{\large EUROPEAN ORGANIZATION FOR NUCLEAR RESEARCH (CERN)}
\vspace*{1.5cm}
\noindent
\begin{tabular*}{\linewidth}{lc@{\extracolsep{\fill}}r@{\extracolsep{0pt}}}
\ifthenelse{\boolean{pdflatex}}
{\vspace*{-1.5cm}\mbox{\!\!\!\includegraphics[width=.14\textwidth]{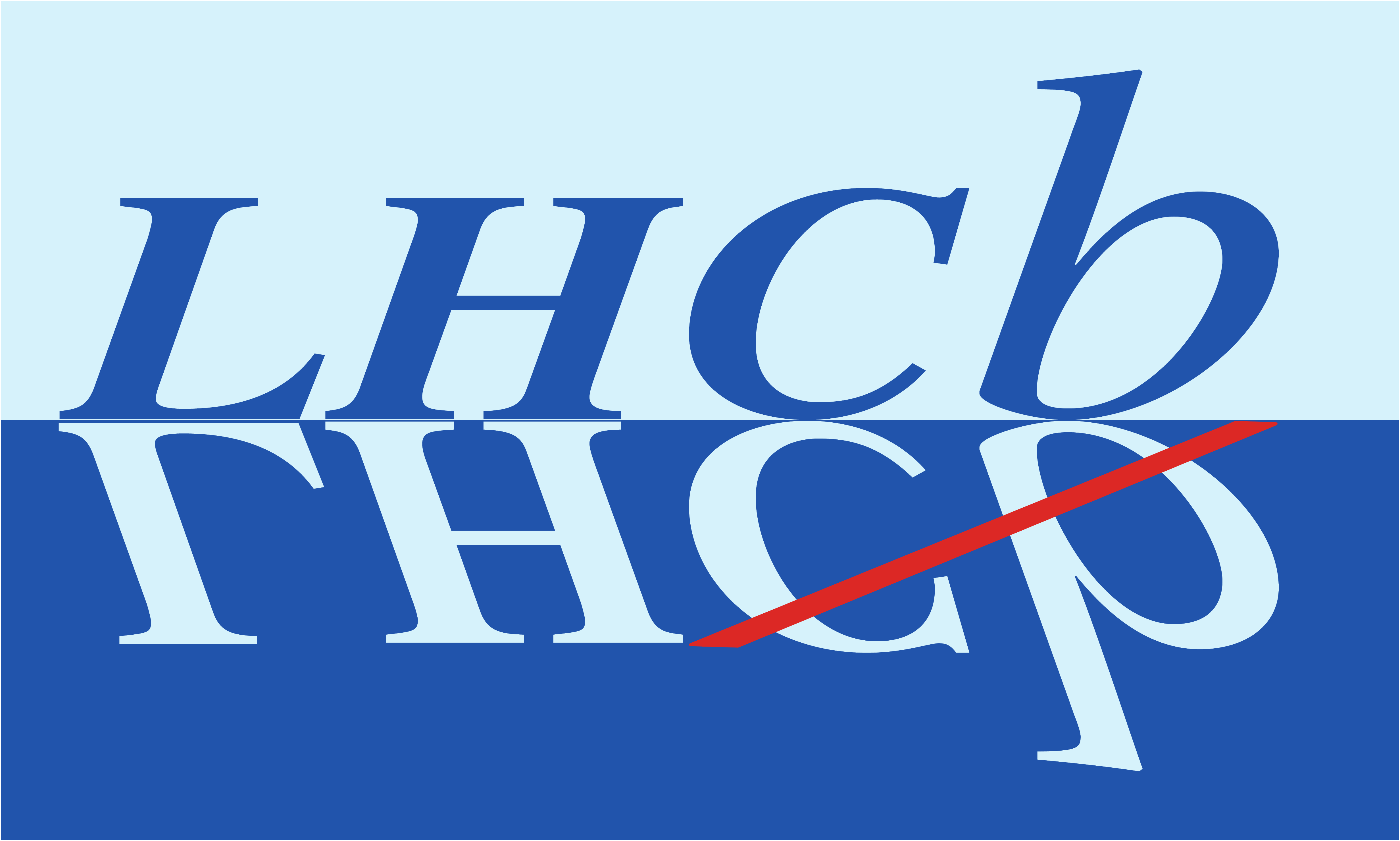}} & &}%
{\vspace*{-1.2cm}\mbox{\!\!\!\includegraphics[width=.12\textwidth]{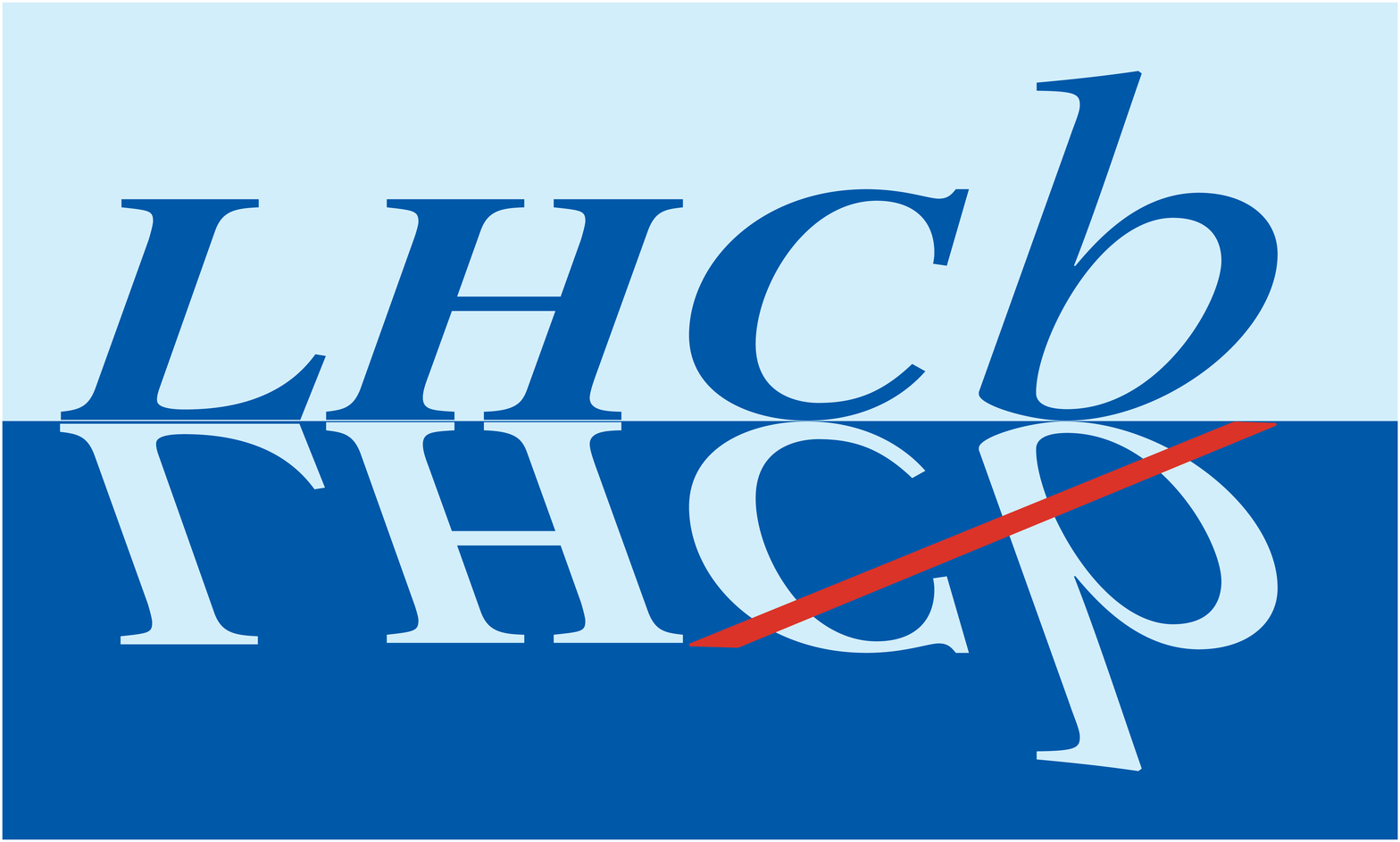}} & &}%
\\
 & & CERN-EP-2022-204 \\  
 & & LHCb-PAPER-2022-034 \\  
 & & \today \\ 
 & & \\
\end{tabular*}

\vspace*{4.0cm}

{\normalfont\bfseries\boldmath\huge
\begin{center}
  \papertitle 
\end{center}
}

\vspace*{2.0cm}

\begin{center}
\paperauthors
\footnote{Authors are listed at the end of this paper.}
\end{center}

\vspace{\fill}

\begin{abstract}
  \noindent
  The ratio of branching fractions of $B_c^+ \to B_s^0 \pi^+$
  and $B_c^+ \to J/\psi \pi^+$ decays is measured with proton-proton collision
  data of a centre-of-mass energy of $13\tev$. The data were collected with the LHCb experiment during 2016--2018,
  corresponding to an integrated luminosity of $5.4\invfb$.
  The $\Bs$ mesons are reconstructed via the decays $\BsToJPsiPhi$ and $\Bs \to \Dsm \pip$.
  The ratio of branching fractions is measured to be $\BR(\Bcp \to \Bs \pip)/\BR(\Bcp \to \jpsi \pip) = 91 \pm 10 \pm 8 \pm 3$
  where 
  the first uncertainty is statistical, the second is systematic
  and the third is due to the knowledge of the branching fractions of the intermediate state decays.

\end{abstract}

\vspace*{2.0cm}

\begin{center}
  Published in JHEP 07 (2023) 066
\end{center}

\vspace{\fill}

{\footnotesize 
\centerline{\copyright~\papercopyright. \href{\paperlicenceurl}{\paperlicence}.}}
\vspace*{2mm}

\end{titlepage}


\newpage
\setcounter{page}{2}
\mbox{~}
%
%
%
%


\renewcommand{\thefootnote}{\arabic{footnote}}
\setcounter{footnote}{0}

\cleardoublepage


\pagestyle{plain} 
\setcounter{page}{1}
\pagenumbering{arabic}


\section{Introduction}%
\label{sec:Introduction}
The $B_c^{(*)+}$ family
of mesons is the only one formed by
two different heavy flavour quarks ($\bquarkbar \cquark$). Both the $\bquarkbar$ quark and the $\cquark$ quark can each
decay with the other as a spectator, 
leading to final states such as $\jpsi\pip$ and $\Bs \pip$. In addition, the $\bquarkbar$ quark and the $\cquark$ quark can 
annihilate via a $\Wp$
boson, allowing pure leptonic final states such as $\taup \neut$.

The $B_c^+ \to B_s^0\pi^+$ decay\footnote{Inclusion of charge conjugate processes is implied throughout this paper.} was first observed in 2013 by the LHCb
collaboration~\cite{LHCb-PAPER-2013-044}, 
and can be used to tag the initial flavour of
the $\Bs$ meson with the charge of the accompanying $\pi^+$ meson.
The $B_{c}^{+}\to B_{s}^{0}\pi^{+}$ decay is also the first observed case of one $B$ meson decaying weakly into another $B$ meson. This property makes the decay ideal for testing theoretical models.
The branching fraction of the $\BsPi$ decay should be large as it is a Cabibbo-favoured decay. 
There are several predictions for the branching fraction of
the $\BsPi$ decay based on QCD sum rules or quark-potential models,
which range between 2.5\%
and 16.4\%~\cite{Kiselev:2000pp, Gouz:2002kk, Ivanov:2006ni,
  Anisimov:1998uk, Colangelo:1999zn, Ebert:2003wc, Dhir:2008zz,
  Naimuddin:2012dy}.
A precise measurement of the branching fraction of the $\BsPi$ decay will improve the
understanding of the $\Bc$ theory models.

The $\Bc \to \taup \neut$ decay is highly sensitive to new physics effects~\cite{Murgui:2019czp, Alonso:2016oyd, Li:2016vvp} but experimental accessibility to this mode is limited. Due to the large branching fraction of the $\BsPi$ decay, its improved measurement contributes to a more stringent limit on the $\Bc \to \taup \neut$ decay via the $\Bc$ total decay width, depending on the theoretical model assumed~\cite{Alonso:2016oyd, Celis:2016azn, Beneke:1996xe}.

In this paper, the ratio between the branching fractions of the
$\BsPi$ and $\JpsiPi$ decays
is measured, using proton-proton ($pp$) collision
data collected with the LHCb experiment between 2016 and 2018 at a centre-of-mass energy of $13\tev$,
corresponding to an integrated luminosity of $5.4\invfb$.
The ratio of branching fractions is measured separately using the $\Bs \to \jpsi(\to \mumu) \phiz(\to\Kp\Km)$ and $\Bs \to \Dsm(\to \Kp\Km\pim) \pip$ decays. These two measurements are then combined.
The ratio of branching fractions is evaluated as
\begin{equation}
\mathcal{R}_{X} \equiv \frac{\BR(\BsPi)}{\BR(\JpsiPi)}
= \frac{N_{\BsPi}}{N_{\JpsiPi}} \cdot \frac{\epsilon_{\JpsiPi}}{\epsilon_{\BsPi}} \cdot \frac{1}{\mathcal{R}_{\rm int}^{X}},
\end{equation}
where $N$ is the signal yield, $\epsilon$ is the total efficiency to reconstruct these decays, and $\mathcal{R}_{\rm int}^{X}$ is the ratio of branching fractions of the corresponding intermediate state decays, as shown in Table~\ref{table:intermediate}.
The $X$ represents the final state of the \Bs decay, $\jpsi\phiz$ or $\Dsm\pip$.
The superscripts or subscripts $\jpsi\phiz$ and $\Dsm\pip$ are used to indicate the $\BsPiJpsiPhi$ and $\BsPiDsPi$ decay modes. The $\mathcal{R}_{\rm int}^{X}$ are defined as
\begin{equation}
    \mathcal{R}_{\mathrm{int}}^{\jpsi \phiz} \equiv \mathcal{B}(\Bs \to \jpsi \phiz)\cdot \mathcal{B}(\phiz \to \Kp \Km),
\end{equation}
\begin{equation}
    \mathcal{R}_{\mathrm{int}}^{\Dsm \pip} \equiv \mathcal{B}(\BsToDsPi)\cdot \mathcal{B}(\DsToKKPi)/\mathcal{B}(\jpsi \to \mumu)
\end{equation}
with corresponding values of  $\mathcal{R}_{\rm int}^{\jpsi \phiz}$ = $(5.11 \pm 0.20)\times 10^{-4}$ and $\mathcal{R}_{\rm int}^{\Dsm \pip}$ = $(2.69\pm 0.14)\times 10^{-3}$.

\begin{table}
  \begin{center}
    \caption{\small \label{table:intermediate}
      Branching fractions of intermediate-state decays~\cite{myPDG2022}. }
      \begin{tabular}{lc}
        \toprule
        Decay & Branching fraction [\%] \\
        \midrule
        $\DsToKKPi$ & $5.38 \pm 0.10$ \\ 
        $\BsToDsPi$ & $0.298 \pm 0.014$ \\ 
        $B_{s}^{0} \to J/\psi\phi$ & $0.104 \pm 0.004$ \\ 
        $\phiz \to K^{+}K^{-}$ & $49.1 \pm 0.5\;\:$ \\
        $\jpsi \to \mumu$ & $5.961 \pm 0.033$ \\
        \bottomrule
      \end{tabular}
  \end{center}
\end{table}

\section{Detector and simulation}%
\label{sec:DetectorAndSimulation}

The \lhcb detector~\cite{LHCb-DP-2008-001,LHCb-DP-2014-002} is a single-arm forward
spectrometer covering the \mbox{pseudorapidity} range $2<\eta <5$,
designed for the study of particles containing \bquark or \cquark
quarks. The detector includes a high-precision tracking system
consisting of a silicon-strip vertex detector surrounding the $pp$
interaction region~\cite{LHCb-DP-2014-001}, a large-area silicon-strip detector located
upstream of a dipole magnet with a bending power of about
$4{\mathrm{\,Tm}}$, and three stations of silicon-strip detectors and straw
drift tubes
placed downstream of the magnet~\cite{LHCb-DP-2013-003}.
The tracking system provides a measurement of the momentum, \ptot, of charged particles with
a relative uncertainty that varies from 0.5\% at low momentum to 1.0\% at 200\gevc.
The minimum distance of a track to a primary $pp$ collision
vertex (PV), the impact parameter (IP),
is measured with a resolution of $(15+29/\pt)\mum$,
where \pt is the component of the momentum transverse to the beam, in\,\gevc.
Different types of charged hadrons are distinguished using information
from two ring-imaging Cherenkov detectors~\cite{LHCb-DP-2012-003}.
Photons, electrons and hadrons are identified by a calorimeter system consisting of
scintillating-pad and preshower detectors, an electromagnetic
and a hadronic calorimeter. Muons are identified by a
system composed of alternating layers of iron and multiwire
proportional chambers~\cite{LHCb-DP-2012-002}.
The online event selection is performed by a trigger~\cite{LHCb-DP-2012-004},
which consists of a hardware stage, based on information from the calorimeter and muon
systems, followed by a software stage, which applies a full event
reconstruction.

Simulation is required to model the effects of the detector acceptance and the
imposed selection requirements. In the simulation, $pp$ collisions are generated using
\pythia8~\cite{Sjostrand:2007gs,*Sjostrand:2006za}
with a specific \lhcb configuration~\cite{LHCb-PROC-2010-056}.
A dedicated generator \bcvegpy~\cite{Chang:2006xka} is used to simulate the production of $\Bc$ mesons.
Decays of unstable particles
are described by \evtgen~\cite{Lange:2001uf}, in which final-state
radiation is generated using \photos~\cite{davidson2015photos}.
The interaction of the generated particles with the detector, and its response,
are implemented using the \geant
toolkit~\cite{Allison:2006ve, *Agostinelli:2002hh} as described in
Ref.~\cite{LHCb-PROC-2011-006}.

\section{Event selection}%
\label{sec:EventSelection}

In the $\Bc \to \Bs \pip$ decay mode, the $\Bs$ meson is reconstructed using two decay modes, $\Bs \to \jpsi \phiz$ and $\Bs \to \Dsm \pip$, with the intermediate states selected based on the ${\DsToKKPi}$, $\jpsi \to \mumu$ and $\phiz \to \Kp \Km$ decays.
The reconstructed $\Bs$ candidates are selected using the boosted decision tree (BDT)~\cite{Breiman,AdaBoost,10.1214/aos/1013203451} classifier 
used in Refs.~\cite{LHCb-PAPER-2013-044,LHCb-PAPER-2020-003}, with the same working points.
In the $\Bc \to \jpsi \pip$ decay channel, the $\jpsi$ mesons are reconstructed using a pair of oppositely charged muons.
The selected $\Bs$ and $\jpsi$ candidates are combined with a track identified as a pion to reconstruct the $\Bc$ candidates.
These candidates and all intermediate states are required to have
good vertex-fit quality. All final state particles (pions, kaons and muons) are required to have a good track-fit quality
and high transverse momentum. Particle identification (PID) information is used to
suppress misidentified tracks.
The reconstructed masses of intermediate state particles are required to be within three times the expected mass resolution of their known masses~\cite{myPDG2022}. 
A second boosted decision tree classifier taken from Refs.~\cite{LHCb-PAPER-2020-003, LHCb-PAPER-2014-050} is used to further distinguish signal $\Bc$ mesons from combinatorial background.
The vertex fit quality, transverse momentum and topological information are used in all BDT classifiers.

In the offline selection, trigger signatures are associated with reconstructed particles.
Selection requirements can therefore be made on the trigger selection itself and on
whether the event was selected at trigger level because of the signal candidate (denoted \verb|TOS|), or because of the decay products of the $b$- or $c$-hadrons produced together with the signal $\Bc$ meson (denoted \verb|TIS|).
For the $\BsPiJpsiPhi$ decay, the reconstructed $\jpsi$ mesons are required
to pass the hardware trigger that selects a muon or a pair of muons with high $\pt$ (the muon \verb|TOS| trigger).
For the $\BsPiDsPi$ decay, hadronic triggers are avoided due to the complexity of simulating the detector response of many-body final states overlapping in the calorimeter. 
Instead, events containing $\BsPi$ decays are required to pass the trigger in case there is a muon or pair of muons in the rest of the event, not related to the $\Bc$ daughter particles (the muon \verb|TIS| trigger). 
This requirement gives better control of trigger efficiencies, even though it reduces the signal yields. 
In the $\RJpsiPhi$ case, the \verb|TOS| trigger requirement is applied to the $\BsPi$ and $\JpsiPi$ decays, and in the 
$\RDsPi$ case, the \verb|TIS| trigger requirement is applied.

\section{Signal yields and efficiencies}%
\label{sec:SignalYieldsAndEfficiency}

\begin{figure}
  \centering
  \subfloat{\includegraphics[scale=0.4]{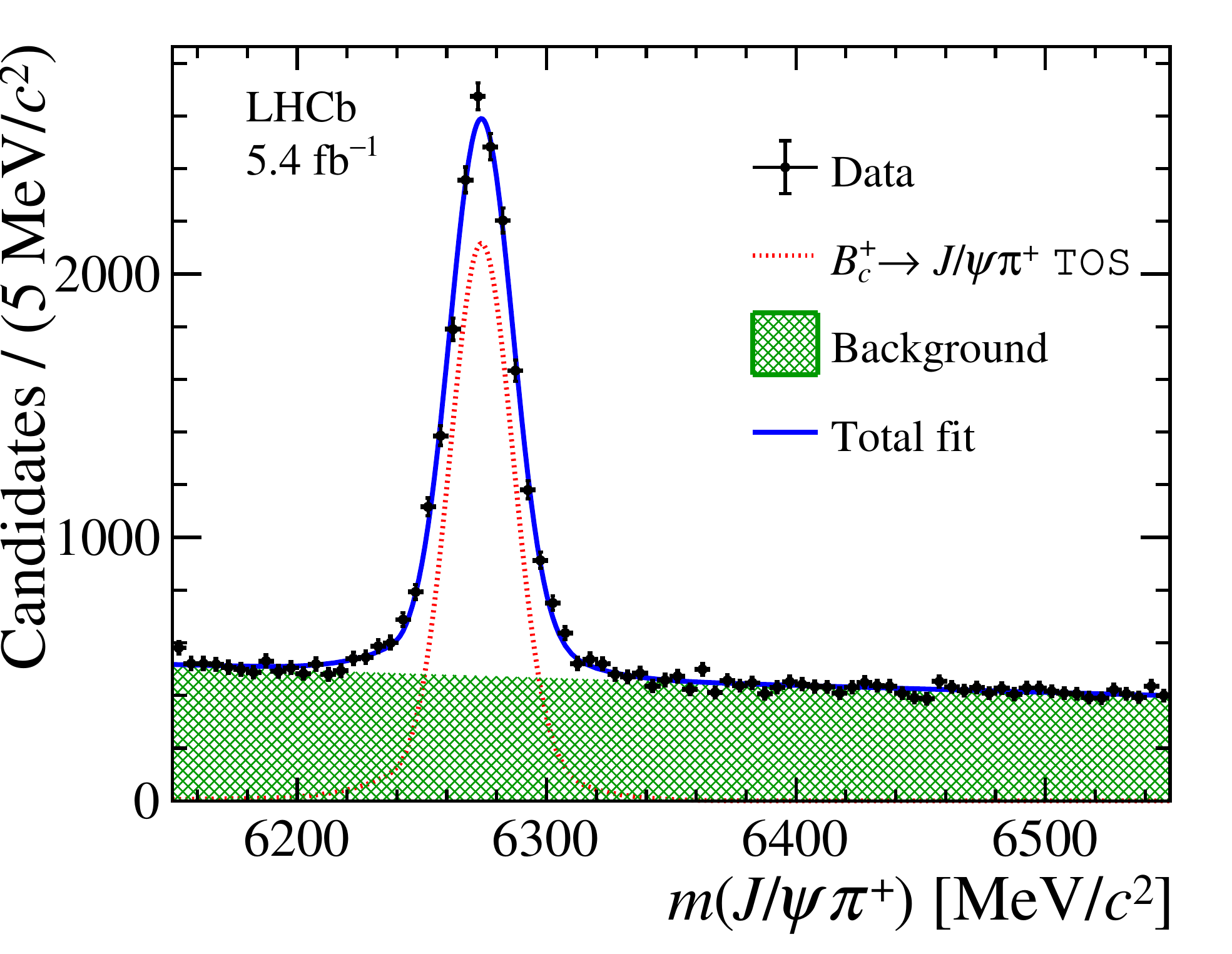}}
  \subfloat{\includegraphics[scale=0.4]{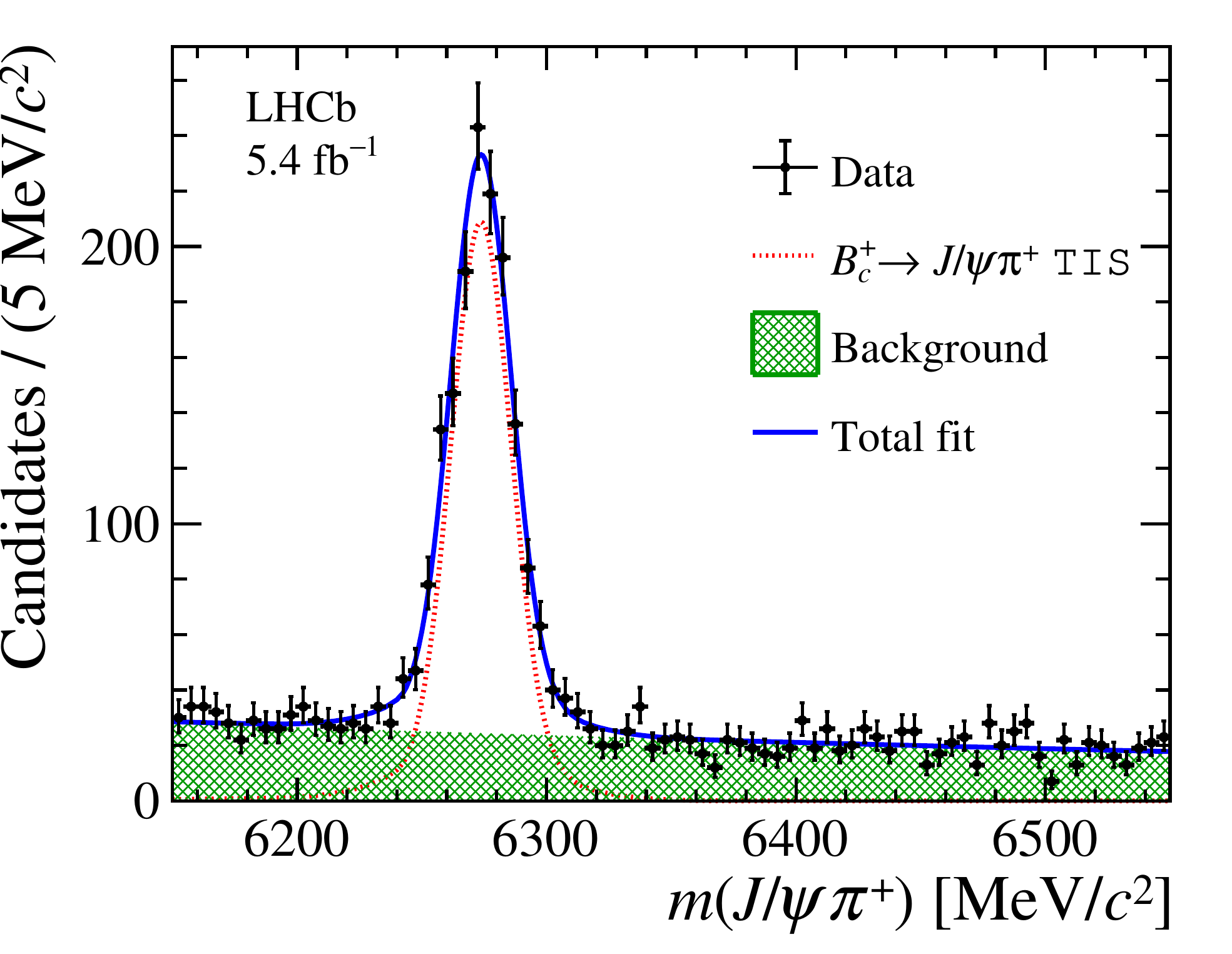}}
  \\
  \subfloat{\includegraphics[scale=0.4]{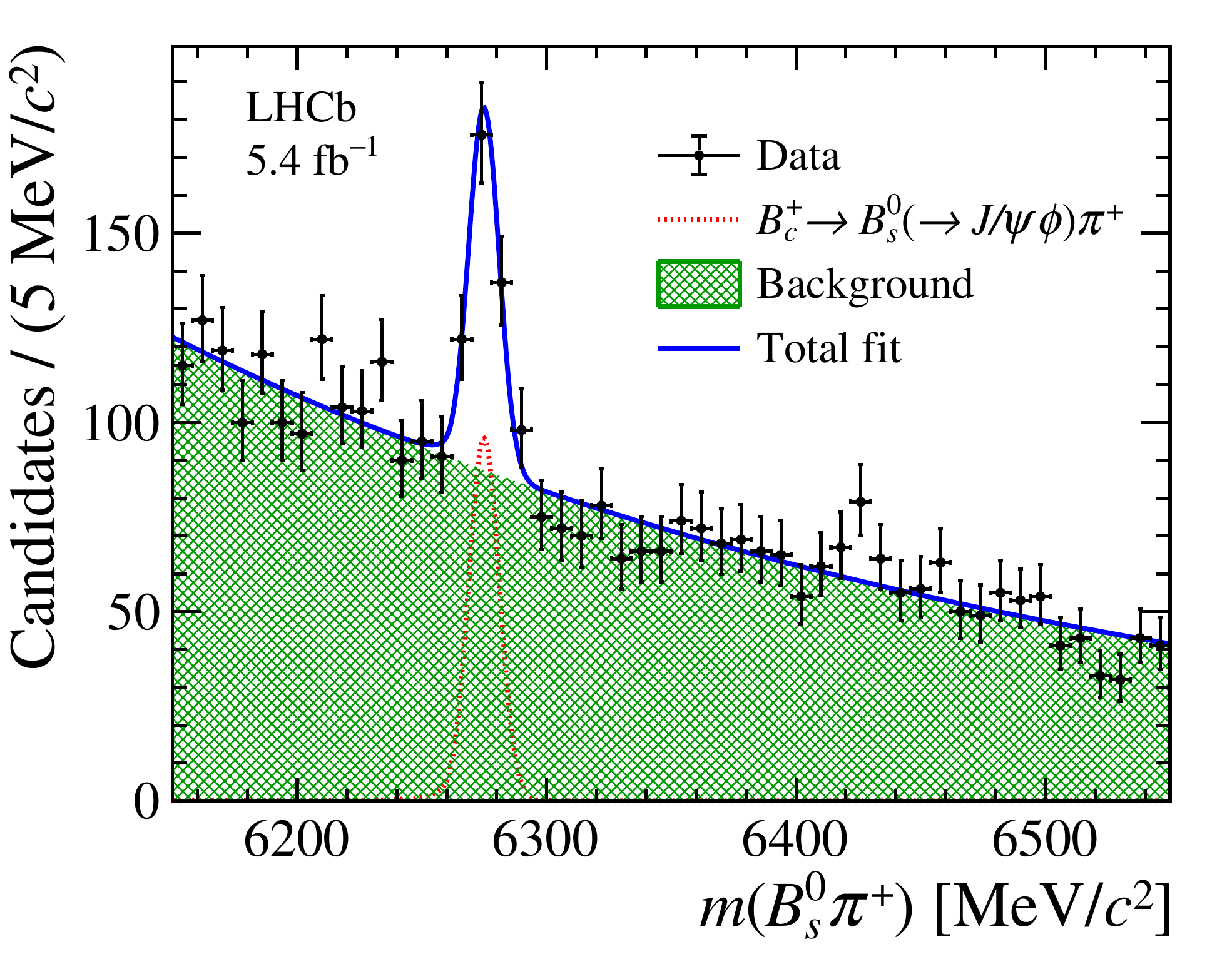}}
  \subfloat{\includegraphics[scale=0.4]{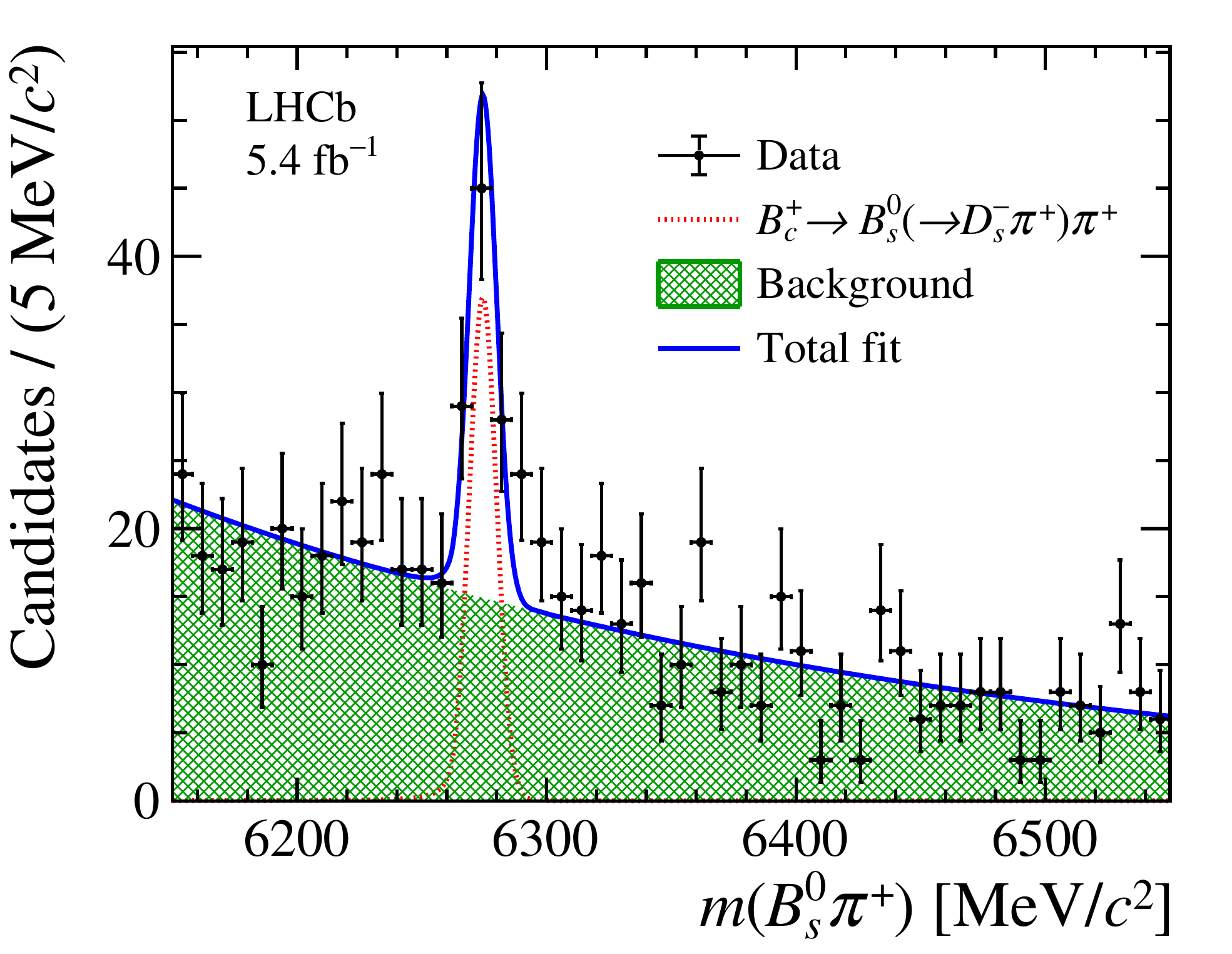}}
  \caption{ \label{fig:fitplots}
    Invariant-mass distributions of $\Bc$ candidates: (top left) $\JpsiPi$ \TOS, (top right) $\JpsiPi$ \TIS, (bottom left) $\BsPiJpsiPhi$, and (bottom right) \mbox{$\BsPiDsPi$}. The black dots represent the data, the blue line represents the results of the total fit,  and the red dashed line and green shaded area show the signal and combinatorial background components, respectively.
    }
\end{figure}

The signal yields are determined from unbinned maximum-likelihood fits to the $\Bc$ candidate invariant-mass distributions.
The invariant mass spectrum of $\Bc$-meson candidates is obtained from a fit to the whole decay chain with the masses of $\Bs$, $\jpsi$, and $\Dsm$ mesons constrained to their known values~\cite{myPDG2022}, as described in Ref.~\cite{Hulsbergen:2005pu}. For the $\Bc \to \jpsi \pip$ decay, the signal component is modelled using the sum of two double-sided Crystal Ball (DSCB) functions~\cite{Skwarnicki:1986xj} with a common peak position. 
For the $\Bc \to \Bs \pip$ decay, one DSCB function is sufficient to model the signal shape. The tail parameters of all 
DSCB functions are determined from simulation, as detailed in Ref.~\cite{LHCb-PAPER-2020-003}.
In both cases the combinatorial background is modelled with an
exponential function. 
The fit results are shown in Fig.~\ref{fig:fitplots}, and the
signal yields are summarised in Table~\ref{tab:signalyields}.

\begin{table}[b]
  \centering
  \caption{\small \label{tab:signalyields}
    Yields of the normalisation and signal decay modes.}
  \begin{tabular}{lr}
    \toprule
    Decay channel & Signal yields\\
    \midrule
    $\Bc \to \jpsi \pip$ (\verb|TOS|) & $14641 \pm 195$ \\
    $\Bc \to \jpsi \pip$ (\verb|TIS|) & $1376 \pm \;\,45$   \\
    $\BsPiJpsiPhi$ & $186 \pm \;\,25$    \\
    $\BsPiDsPi$   & $65 \pm \;\,12$     \\
    \bottomrule
  \end{tabular}
\end{table}

The total efficiency is calculated as the product of the geometrical detector acceptance
and of the efficiencies related to particle reconstruction, event selection, particle identification and trigger decision. 
The efficiencies are determined using simulation calibrated with data. 
Since the number of final-state particles is different between the signal and normalisation modes (five and three, respectively), 
accurate modelling of the track reconstruction efficiency is important. For every event, a track-by-track correction to the track reconstruction efficiency of all final states is applied
using a calibration sample of $\jpsi \to \mumu$ decays~\cite{LHCb-DP-2013-002}.
The PID information used in the event selection is corrected with high-yield calibration
samples ($\Dstarp \to \Dz\pip$ with $\Dz \to \Km \pip$ decays for hadrons, and the
$\jpsi \to \mumu$ decay for muons)~\cite{LHCb-DP-2018-001}. The total efficiencies are listed in Table~\ref{tab:eff}.

\begin{table}
  \centering
  \caption{\small \label{tab:eff}
    Total efficiencies of the $\JpsiPi$ and $\BsPi$ decay modes. The uncertainties are due to the finite size of the simulated samples.}
  \begin{tabular}{lc}
    \toprule
    Decay channel & Efficiency [\%]\\
    \midrule
    $\Bc \to \jpsi \pip$ (\verb|TOS|) & $1.6141 \pm 0.0028$\\
    $\Bc \to \jpsi \pip$ (\verb|TIS|) & $0.15846 \pm 0.00063$\\
    $\BsPiJpsiPhi$ & $0.4526 \pm 0.0013$\\
    $\BsPiDsPi$  & $0.02801 \pm 0.00028$\\
    \bottomrule
  \end{tabular}
\end{table}

\section{Systematic uncertainties}%
\label{sec:SystematicUncertainties}
The measurement is affected by the systematic uncertainty in the determination of signal yields and efficiencies, as summarised in Table~\ref{tab:susum}.

The systematic uncertainty due to the signal lineshape modelling is studied 
using pseudoexperiments. 
The signal candidates in the fully simulated samples are mixed with background events that are randomly generated with the exponential shape and fraction determined from the fits to the data. 
The same fit model used for the data fit is applied to these samples. 
The difference between the fitted value of the signal yield ratio $N_{\BsPi}/N_{\JpsiPi}$ and the generated ratio in simulated samples is taken as the systematic uncertainty, 3.0\% for $\RJpsiPhi$ and 4.4\% for $\RDsPi$.   
The systematic uncertainty of the background lineshape is estimated by using a first-order polynomial function as an alternative shape, and the difference with the default result is taken as the systematic uncertainty, which is 5.5\% for $\RJpsiPhi$ and 4.5\% for $\RDsPi$.

The $\Bc$ mesons can also decay to a $B_{s}^{*0}\pi^{+}$ state, followed by the $B_{s}^{*0} \to
\Bs\gamma$ decay. This final state, with an unobserved photon, mimics the signal but with the reconstructed $m(\Bs\pip)$ shifted to lower values by the mass difference between $B_{s}^{*0}$ and $\Bs$ mesons to a good approximation. In this case, the $\Bc$ mass resolution is almost
unaffected~\cite{LHCb-PAPER-2013-044, LHCb-PAPER-2019-007}.
To estimate the impact of this potential partially reconstructed decay, an alternative fit to the $m(\Bs\pip)$ mass distribution is performed including an additional $B_{s}^{*0}\pi^{+}$ component. 
The fit model for $\Bc\to B_{s}^{*0}\pi^{+}$ is a DSCB function with
the same tail parameters as in the default fit. The mass resolution is assumed to be the same as the signal
and the mass shift is fixed to the known mass difference between $\Bs$ and $B_{s}^{*0}$ mesons, $48.5^{+1.8}_{-1.5}\mevcc$~\cite{myPDG2022}. The difference in the signal yield ratio is taken as the relevant systematic uncertainty, which is $3.2\%$ for $\RJpsiPhi$ and $3.1\%$ for $\RDsPi$.

As some efficiencies are determined from simulated samples, any discrepancy between data and simulated events can introduce a bias. 
Distributions of kinematic variables in simulation and in background-subtracted data using the {\it sPlot} technique~\cite{Pivk:2004ty} are compared. The differences of the efficiencies between the unchanged simulation and after the alignment between data and simulation are 2.3\% for $\RJpsiPhi$ and 4.8\% for $\RDsPi$, respectively, which are taken as systematic uncertainties.

The tracking efficiency, which is determined on data calibration samples, has a systematic uncertainty of 1\% per track.
Due to the hadronic interactions and the uncertainty of the LHCb material-budget in simulation, there is 1.1\% additional uncertainty for kaons and 1.4\% additional uncertainty for pions.
The uncertainties for the same particles in numerator and denominator largely cancel in the ratio. Therefore, the systematic uncertainty is found to be 3.2\% for the $\RJpsiPhi$ measurement and 6.0\% for $\RDsPi$.

The efficiencies determined from simulated events have uncertainties due to the limited size of the samples. The relative uncertainty in the total efficiency ratio $\epsilon (\JpsiPi)/ \epsilon (\BsPi)$ is considered as the systematic uncertainty, the value is 0.3\% for $\RJpsiPhi$ and 1.1\% for $\RDsPi$. 
The imperfect simulation of the trigger system can also cause a bias in the trigger selection efficiency. An uncertainty of 1.0\% is assigned to the trigger selection following Ref.~\cite{LHCb-PAPER-2016-057}. 
Other potential sources of systematic uncertainty, such as those associated with the PID calibration, are found to be negligible.
A summary of the systematic uncertainties is given in Table~\ref{tab:susum}.

\begin{table}
    \begin{center}
        \caption{\small \label{tab:susum}
          Relative systematic uncertainties in the ratio of the branching fractions where the symbol $\RJpsiPhi$ indicates the ratio of branching fractions measured with the $\BsPiJpsiPhi$ decay channel and $\RDsPi$ that using the $\BsPiDsPi$ decay. }
        \begin{tabular}{lcc}
            \toprule
            & \RJpsiPhi~[\%] & \RDsPi~[\%] \\
            \midrule
            Signal model & 3.0 &4.4  \\
            Background model & 5.5 & 4.5  \\
            $\Bc \to B_{s}^{*0}\pip$ component & 3.2 & 3.1 \\
            Data-simulation agreement      & 2.3 & 4.8 \\
            Tracking efficiency & 3.2 & 6.0 \\
            Simulation sample size &  0.3 & 1.1 \\
            Trigger & 1.0 & 1.0 \\
            \midrule
            Total & 8.1 & 10.5\;\, \\
            \bottomrule
        \end{tabular}
    \end{center}
\end{table}

\section{Results}%
\label{sec:Result}

The ratio of the total branching fractions, including the branching fractions of the
intermediate states, is determined as
\begin{equation}
\RJpsiPhi^{\mathrm{tot}} \equiv \frac{N_{\BsPiJpsiPhi}}{N_{\JpsiPi}} \cdot \frac{\epsilon_{\JpsiPi}}{\epsilon_{\BsPiJpsiPhi}},
\label{R1}
\end{equation}
\begin{equation}
\RDsPi^{\mathrm{tot}} \equiv \frac{N_{\BsPiDsPi}}{N_{\JpsiPi}}
\cdot \frac{\epsilon_{\JpsiPi}}{\epsilon_{\BsPiDsPi}},
\end{equation}
and are measured to be
\begin{equation*}
  \RJpsiPhi^{\mathrm{tot}} =0.0453 \pm 0.0061\stat\pm 0.0037\syst,
\end{equation*}
\begin{equation*}
\RDsPi^{\mathrm{tot}} =0.267 \pm 0.050 \stat \pm 0.028\syst,
\end{equation*}
in each decay mode. Accounting for the branching fractions
of the intermediate state decays, as shown in Table~\ref{table:intermediate}, the ratio between the branching fractions of the
$\BsPi$ and $\JpsiPi$ decays is
\begin{equation*}
\RJpsiPhi = 89 \pm 12\stat \pm 7 \syst \pm 4\,(\mathcal{B}),
\end{equation*}
\begin{equation*}
\RDsPi = 99 \pm 19 \stat \pm 10 \syst \pm 5\, (\mathcal{B}),
\end{equation*}
for the two decay modes considered. The third uncertainty is due to the knowledge of the branching fractions of intermediate state decays (denoted $\mathcal{B}$ hereafter).

The ratio of the branching fractions  measured using the two decay modes can then be combined using the BLUE
method~\cite{Valassi:2003mu}. The systematic
uncertainties due to the signal lineshape, background lineshape, $\Bc \to B_{s}^{*0}\pip$ component, data-simulation agreement,
tracking efficiency and trigger decision are considered to
be 100\% correlated between the two decay modes, while the uncertainty due to limited simulation sample size is uncorrelated. Uncertainties in this ratio arise from knowledge of the branching fractions of intermediate state decays, which is the uncertainty of $\mathcal{R}_{\rm int}$. The uncertainties from $\mathcal{R}_{\rm int}^{\jpsi \phiz}$ and $\mathcal{R}_{\rm int}^{\Dsm \pip}$ are considered uncorrelated.
The combined result is
\begin{equation*}
\mathcal{R} = \frac{\BR(\BsPi)}{\BR(\JpsiPi)} = 91 \pm 10\stat \pm 8 \syst \pm 3\,(\mathcal{B}).
\end{equation*}
This is the first direct measurement of the ratio of branching fractions \mbox{$\BR(\Bcp \to \Bs \pip)/\BR(\Bcp \to \jpsi \pip)$}.

\section{Discussion of $\boldsymbol{\mathcal{B}(\Bc \to \Bs \pip)}$}
The branching fraction of the $\Bc \to \Bs \pip$ decay can be extracted from this result with the knowledge of $\mathcal{B}(\JpsiPi)$.
The branching fraction of the $\JpsiPi$ decay is predicted to be $(0.291^{+0.043}_{-0.050})\%$~\cite{Qiao:2012hp} with a $\Bc$ lifetime of 0.453\ps. Using this input and the $\Bc$ lifetime of 0.510\ps taken from Ref.~\cite{myPDG2022}, the branching fraction of the $\Bc \to \Bs \pip$ decay is found to be
\begin{equation}
\BR(\BsPi) = \left(30.0 \pm 3.3 \stat \pm 2.6 \syst \pm 1.0\,(\mathcal{B})^{+4.6}_{-5.5}\,(\JpsiPi)\right)\%,
\label{result1}
\end{equation}
where the fourth uncertainty is the uncertainty from the prediction of $\mathcal{B}(\JpsiPi)$.

The branching fraction of the $\JpsiPi$ decay can also be determined from the theoretical prediction of $\mathcal{B}(\Bc \to \jpsi \mup \neum)$
and the measurement of \mbox{$\mathcal{B}(\JpsiPi)/\mathcal{B}(\Bcp \to \jpsi \mup \neum)= 0.0469 \pm 0.0028 \pm 0.0046$}~\cite{LHCB-PAPER-2014-025}.
The average of several theoretical predictions for $\mathcal{B}(\Bc \to \jpsi \mup \neum)$, as considered in  Ref.~\cite{LHCb-PAPER-2019-033}, is $\mathcal{B}(\Bc \to \jpsi \mup \neum) = (1.95 \pm  0.46) \%$. Using these inputs, the value of $\mathcal{B}(\JpsiPi)$ is found to be $(0.091 \pm 0.024)\%$. The branching fraction of the $\BsPi$ decay is determined to be
\begin{equation}
  \BR(\BsPi) = (8.3 \pm 0.9 \stat \pm 0.7 \syst \pm 0.3(\mathcal{B}) \pm 2.2\,(\JpsiPi))\%.
  \label{result2}
\end{equation}
The fourth uncertainty in Eq.~\ref{result2} is due to the uncertainty on the value of $\mathcal{B}(\JpsiPi)$.

Equations \ref{result1} and \ref{result2} show a discrepancy in  $\mathcal{B}(\Bc \to \Bs \pip)$ which is larger than the quoted uncertainties, due to different predictions for the value of $\mathcal{B}(\JpsiPi)$.
Until this discrepancy is resolved, caution is advised when converting the measured ratio into a measurement of $\mathcal{B}(\Bc \to \Bs \pip)$. 
However, irrespective of this discrepancy, either value of the branching fraction $\mathcal{B}(\JpsiPi)$ considered here results in the $\Bc \to \Bs \pip$ decay having the largest branching fraction of all $\Bcp$ decays measured to date.

\section*{Acknowledgements}
%
%
\noindent We express our gratitude to our colleagues in the CERN
accelerator departments for the excellent performance of the LHC. We
thank the technical and administrative staff at the LHCb
institutes.
We acknowledge support from CERN and from the national agencies:
CAPES, CNPq, FAPERJ and FINEP (Brazil); 
MOST and NSFC (China); 
CNRS/IN2P3 (France); 
BMBF, DFG and MPG (Germany); 
INFN (Italy); 
NWO (Netherlands); 
MNiSW and NCN (Poland); 
MEN/IFA (Romania); 
MICINN (Spain); 
SNSF and SER (Switzerland); 
NASU (Ukraine); 
STFC (United Kingdom); 
DOE NP and NSF (USA).
We acknowledge the computing resources that are provided by CERN, IN2P3
(France), KIT and DESY (Germany), INFN (Italy), SURF (Netherlands),
PIC (Spain), GridPP (United Kingdom), 
CSCS (Switzerland), IFIN-HH (Romania), CBPF (Brazil),
Polish WLCG  (Poland) and NERSC (USA).
We are indebted to the communities behind the multiple open-source
software packages on which we depend.
Individual groups or members have received support from
ARC and ARDC (Australia);
Minciencias (Colombia);
AvH Foundation (Germany);
EPLANET, Marie Sk\l{}odowska-Curie Actions and ERC (European Union);
A*MIDEX, ANR, IPhU and Labex P2IO, and R\'{e}gion Auvergne-Rh\^{o}ne-Alpes (France);
Key Research Program of Frontier Sciences of CAS, CAS PIFI, CAS CCEPP, 
Fundamental Research Funds for the Central Universities, 
and Sci. \& Tech. Program of Guangzhou (China);
GVA, XuntaGal, GENCAT and Prog.~Atracci\'on Talento, CM (Spain);
SRC (Sweden);
the Leverhulme Trust, the Royal Society
 and UKRI (United Kingdom).




\addcontentsline{toc}{section}{References}
\bibliographystyle{LHCb}
\bibliography{main,BcReview,standard,LHCb-PAPER,LHCb-CONF,LHCb-DP,LHCb-TDR}

\newpage
\centerline
{\large\bf LHCb collaboration}
\begin
{flushleft}
\small
R.~Aaij$^{32}$\lhcborcid{0000-0003-0533-1952},
A.S.W.~Abdelmotteleb$^{50}$\lhcborcid{0000-0001-7905-0542},
C.~Abellan~Beteta$^{44}$,
F.~Abudin{\'e}n$^{50}$\lhcborcid{0000-0002-6737-3528},
T.~Ackernley$^{54}$\lhcborcid{0000-0002-5951-3498},
B.~Adeva$^{40}$\lhcborcid{0000-0001-9756-3712},
M.~Adinolfi$^{48}$\lhcborcid{0000-0002-1326-1264},
P.~Adlarson$^{77}$\lhcborcid{0000-0001-6280-3851},
H.~Afsharnia$^{9}$,
C.~Agapopoulou$^{13}$\lhcborcid{0000-0002-2368-0147},
C.A.~Aidala$^{78}$\lhcborcid{0000-0001-9540-4988}
% ...

{\footnotesize \it

% ...
$^{9}$Universit{\'e} Clermont Auvergne, CNRS/IN2P3, LPC, Clermont-Ferrand, France\\
% ...
$^{13}$LPNHE, Sorbonne Universit{\'e}, Paris Diderot Sorbonne Paris Cit{\'e}, CNRS/IN2P3, Paris, France\\
% ...
$^{32}$Nikhef National Institute for Subatomic Physics, Amsterdam, Netherlands\\
% ...
$^{40}$Instituto Galego de F{\'\i}sica de Altas Enerx{\'\i}as (IGFAE), Universidade de Santiago de Compostela, Santiago de Compostela, Spain\\
% ...
$^{44}$Physik-Institut, Universit{\"a}t Z{\"u}rich, Z{\"u}rich, Switzerland\\
% ...
$^{48}$H.H. Wills Physics Laboratory, University of Bristol, Bristol, United Kingdom\\
% ...
$^{50}$Department of Physics, University of Warwick, Coventry, United Kingdom\\
% ...
$^{53}$School of Physics and Astronomy, University of Glasgow, Glasgow, United Kingdom\\
$^{54}$Oliver Lodge Laboratory, University of Liverpool, Liverpool, United Kingdom\\
% ...
$^{62}$Syracuse University, Syracuse, NY, United States\\
% ...
$^{77}$Department of Physics and Astronomy, Uppsala University, Uppsala, Sweden, associated to $^{53}$\\
$^{78}$University of Michigan, Ann Arbor, MI, United States, associated to $^{62}$\\
\bigskip
% ...
}
\end{flushleft}

\end{document}